\pdfoutput=1

\documentclass[11pt]{article}

\usepackage[preprint]{coling}

\usepackage{times}
\usepackage{latexsym}
\usepackage{times}
\usepackage{graphicx}
\usepackage{amssymb}
\usepackage{multirow}
\usepackage{booktabs,arydshln}
\usepackage{dcolumn}
\usepackage{array}
\usepackage{tabularx}
\usepackage{xcolor}
\usepackage{comment}
\usepackage{tablefootnote}
\usepackage{enumitem}
\usepackage{threeparttable}
\usepackage{cleveref}
\crefname{subsection}{subsection}{subsections}

\usepackage[T1]{fontenc}

\usepackage[utf8]{inputenc}

\usepackage{microtype}

\usepackage{inconsolata}

\usepackage{graphicx}

\title{AI-powered Digital Framework for Personalized Economical Quality Learning at Scale}

\author{Mrzieh VatandoustMohammadieh, Mohammad Mahdi Mohajeri, Ali Keramati, \and Majid Nili Ahmadabadi \\
        University of Tehran \\
        m.vatandoust@ut.ac.ir \\ mehdimohajeri@ut.ac.ir \\ alikeramati@ut.ac.ir \\ mnili@ut.ac.ir}

\begin{document}
\maketitle
\begin{abstract}
The disparity in access to quality education is significant, both between developed and developing countries and within nations, regardless of their economic status. Socioeconomic barriers and rapid changes in the job market further intensify this issue, highlighting the need for innovative solutions that can deliver quality education at scale and low cost. 
This paper addresses these challenges by proposing an AI-powered digital learning framework grounded in Deep Learning (DL) theory. The DL theory emphasizes learner agency and redefines the role of teachers as facilitators, making it particularly suitable for scalable educational environments. We outline eight key principles derived from learning science and AI that are essential for implementing DL-based Digital Learning Environments (DLEs). Our proposed framework leverages AI for learner modelling based on Open Learner Modeling (OLM), activity suggestions, and AI-assisted support for both learners and facilitators, fostering collaborative and engaging learning experiences. Our framework provides a promising direction for scalable, high-quality education globally, offering practical solutions to some of the AI-related challenges in education.
\end{abstract}

\section{Introduction}
\footnotetext{The term "StudyChum" stems from a traditional practice in Iranian schools, where the teacher briefly introduces a topic, and learners then engage in peer discussions to grasp the concept. The individual guiding the learner through this process is called a "StudyChum" in Persian.}

The gap in access to quality education is already very wide within developed and developing countries \cite{meyer2024reducing, unicefBillionChildren}. This gap is also wide inside each country, regardless of its GDP. Nevertheless, the gap is much deeper in developing countries. Providing quality education at scale and low cost is the key solution to this problem. Gaining access to quality education is among the main reasons for immigration within and between countries, which causes economic, social, cultural, and environmental problems for all. Therefore, providing quality education all over the world is a major need.

All said aside, the gap in access to quality education is widening worldwide due to fast changes in the job market, as well as technological and cultural changes. Disruptive technologies fuel the restructuring of the job market the most. Adaptation to the job market necessitates reskilling and upskilling at a moderate frequency, affordable duration, and cost at scale. To remain relevant and tackle real-world challenges, individuals must continuously acquire in-demand skills, including analytical thinking, technological literacy, flexibility, and lifelong learning (LLL) – a self-driven pursuit of knowledge and adaptability that empowers them to navigate the evolving demands of work and society \cite{UIL2023,weforumFutureJobs}.

Quality education these days is defined in three but tightly bounded dimensions: knowledge, expertise, and soft skills. To narrow the gap in access to quality education, improving soft skills like critical thinking, analytical thinking, teamwork, and communication should be embedded in the learning environment \cite{weforumFutureJobs}. Access to well-educated and experienced teachers and well-formed learning environments and activities are the basics for quality education regarding knowledge, expertise, and soft skills. 

\begin{figure}[ht]
\centerline{\includegraphics[width=230pt]{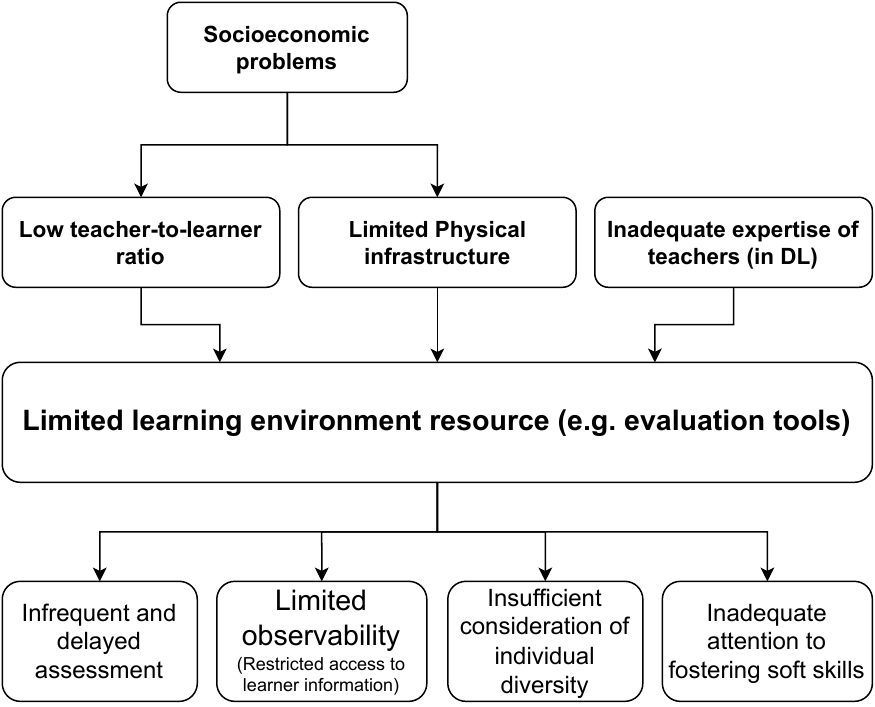}}
 \caption{Categorizing challenges for integrating innovative learning strategies in traditional settings}
 \label{fig:challenges}
\end{figure}

Nevertheless, socioeconomic problems are the main barriers to providing these basics at scale and low cost. These problems are often faced in underprivileged areas, resulting in low teacher-to-learner ratios and inadequate infrastructure, which ultimately constrains the learning environment. Also, a lack of professional educators impedes progress in providing LLL opportunities for all \cite{UIL2023}. These limitations have cascading effects (See \Cref{fig:challenges}).
Resource limitations may lead to less frequent and delayed assessments and impact continuous evaluation. Furthermore, traditional assessment tools, like standardized tests, often capture a low-frequency snapshot of the learning process, leaving teachers mostly unaware of the learners' learning process. This limited observability of learners hinders the gathering of comprehensive information, potentially leading to issues with differentiation and personalization of instruction \cite{hussein2020pedagogical}.
Consequently, teachers may struggle to accurately observe all learners' progress and cater to individual learning styles and paces. This often results in a one-size-fits-all instructional approach, harming learner engagement and achievement.  Furthermore, effectively instructing soft skills, such as communication, collaboration, and critical thinking, can be challenging within a traditional, teacher-centred framework.

Fortunately, wide coverage of the internet and access to smartphones, as well as fast-growing familiarity with using digital tools at all ages, have provided a golden opportunity, especially for developing countries, to exploit digital education technologies for the creation of rich digital learning environments (DLEs), as a complementary and in parallel to face-to-face ones, for quality education at scale and low cost. Nevertheless, successful experiences in DLEs, in terms of efficacy, are very limited, and their scalability at low cost has yet to be studied.

To develop soft skills and attain knowledge and expertise, the target learning environment should be a collaborative and proactive environment that offers personalized learning experiences. DLEs must evolve beyond simple content delivery to embrace interactive and collaborative learning experiences. A passive, one-directional approach with DLEs fails to capitalize on their ability to personalize learning and promote deeper understanding \cite{MFBD2018, CKPY2021}. 
To construct an effective learning environment and associated activities, we evaluated various learning science theories that support quality education at scale. Deep Learning (DL) emerged as the most suitable theory for our objectives. DL's emphasis on developing soft skills and promoting learner agency aligns closely with our goals and makes it particularly well-suited for large-scale implementation. We present a discussion of DL in \Cref{sec:DL}.

We firmly believe that the presence of experienced teachers as "facilitators" in DLEs is highly necessary. Their role is crucial in providing opportunities for enhancing learners' knowledge and developing their soft skills. Additionally, facilitators should provide immediate feedback to learners, which is essential for effective learning. However, we face the bottleneck of limited access to experienced teachers at scale \cite{meyer2024reducing}. To mitigate this issue, we present a digital learning framework in which Artificial Intelligence (AI) reduces the workload of facilitators and supports the DL guidelines.

Despite the potential, most contemporary AI tools and methods have not been developed for human-in-the-loop applications in general \cite{holmberg2020feature} and for educational purposes in particular \cite{america2023doc}. Additionally, AI tools not specifically tailored for education may harm long-term learning, potentially hindering learners' ability to develop critical skills independently \cite{bastani2024generative}. Therefore, in this paper, we mention the existing challenges of using AI in our framework and suggest some directions for solutions.
In light of this, our paper makes the following key contributions:
\begin{enumerate}
    \item {\textbf{Design Principles}: We propose a set of principles derived from DL theory for designing and developing AI-powered DLEs to facilitate high-quality, scalable, and cost-effective educational experiences.}
    \item{\textbf{AI-powered Digital Learning Framework}: We propose an AI-powered digital learning framework based on our proposed design principles in a learning-science-informed manner. This framework enables proactive and personalized learning experiences by embedding in-demand AI and technological tools in education while enhancing essential soft skills attainable within DLEs.}
    \item{\textbf{Addressing AI Challenges in Framework with Practical Solutions:} We identify key AI challenges in our framework and propose practical solutions to mitigate some of these challenges, providing insights for future implementations and research.}
\end{enumerate}

\section{Deep Learning Potentials for Low-cost Quality Education at Scale}
\label{sec:DL}

DL shifts away from traditional education and emphasizes mastery and conceptual understanding over rote memorization. DL can foster lifelong learners' development by equipping them with the requisite skills and mindset for continuous learning and growth. DL connects learning with real-world problems, and mastering core content happens through doing a set of learning activities that inherently facilitate problem-solving, critical thinking, communication, collaboration, creativity, and learning how to learn \cite{NRC2012, hewlett2013, partner21cent2011, fullan2017deep}. Therefore, DL is effective for LLL and soft skills development due to its emphasis on active engagement and metacognitive skills.

DL underscores the learner's central role in the learning process, positioning the learner as an active agent central to knowledge acquisition \cite{fullan2017deep}. It empowers learners to take charge of their education, reducing reliance on direct instruction. Concurrently, DL redefines the role of the teacher as a facilitator. Altogether, this theory is potentially a suitable pedagogical approach for large-scale implementation in the face of limited or lack of access to experienced facilitators. In other words, DL's learner-centred approach makes it suitable for scaling education, especially in areas with low teacher-to-learner ratios. Nonetheless, designing a conducive and low-cost learning environment that facilitates the emergence and advancement of DL continues to be a critical challenge.

As mentioned, for DL to occur, each learner should have an agency for learning. Such an agency emerges when the learning environment and activities are tuned for personalization in a closed loop and semi-real-time manner. Personalization requires learning-science-informed continual learner modelling and a rich set of tools and learning activities \cite{florian2011activity}. Such requirements cannot be provided in physical environments at low cost, with limited access to experienced facilitators and in populated classrooms. Here, we suggest our AI-enabled DLE framework to solve this core problem.

The proposed framework incorporates AI to support DL principles and enhance learning environments at scale and low cost by taking three roles. The first role is to augment facilitators' capabilities by learner modelling and suggesting some personalized learning activities based on learner activity data as a decision-support system for facilitators. The second role of AI is to provide assistance to facilitators using a well-prompted LLM. The third role is providing a personalized learning partner and teaching assistance for learners using a personalized prompted and RAG-enriched LLM. 

\section{DL-based Framework for Scalable AI-Powered Personalized Learning} \label{sec:framework}

As mentioned in \Cref{sec:DL}, the learning environment's richness, particularly in terms of learner modelling and diverse learning activities, is crucial for effective personalization and, consequently, creating agency in the learners. Obviously, diversity in the learning activities is necessary to cover the naturally high diversity in learners' backgrounds, expertise, interests, characteristics, learning styles, etc. Nevertheless, it is also crucial for learner modelling because learners exhibit varying levels of engagement, aptitude and skills across different learning activities.  As a result, learners must be engaged and monitored in different learning tasks at proper frequencies. In other words, creating a closed loop persistent activation of all learning participants— both facilitators and learners—is crucial for consistently augmenting learners' agency \cite{fullan2017deep}.

\begin{figure*}[ht]
\centerline{\includegraphics[width=\textwidth]{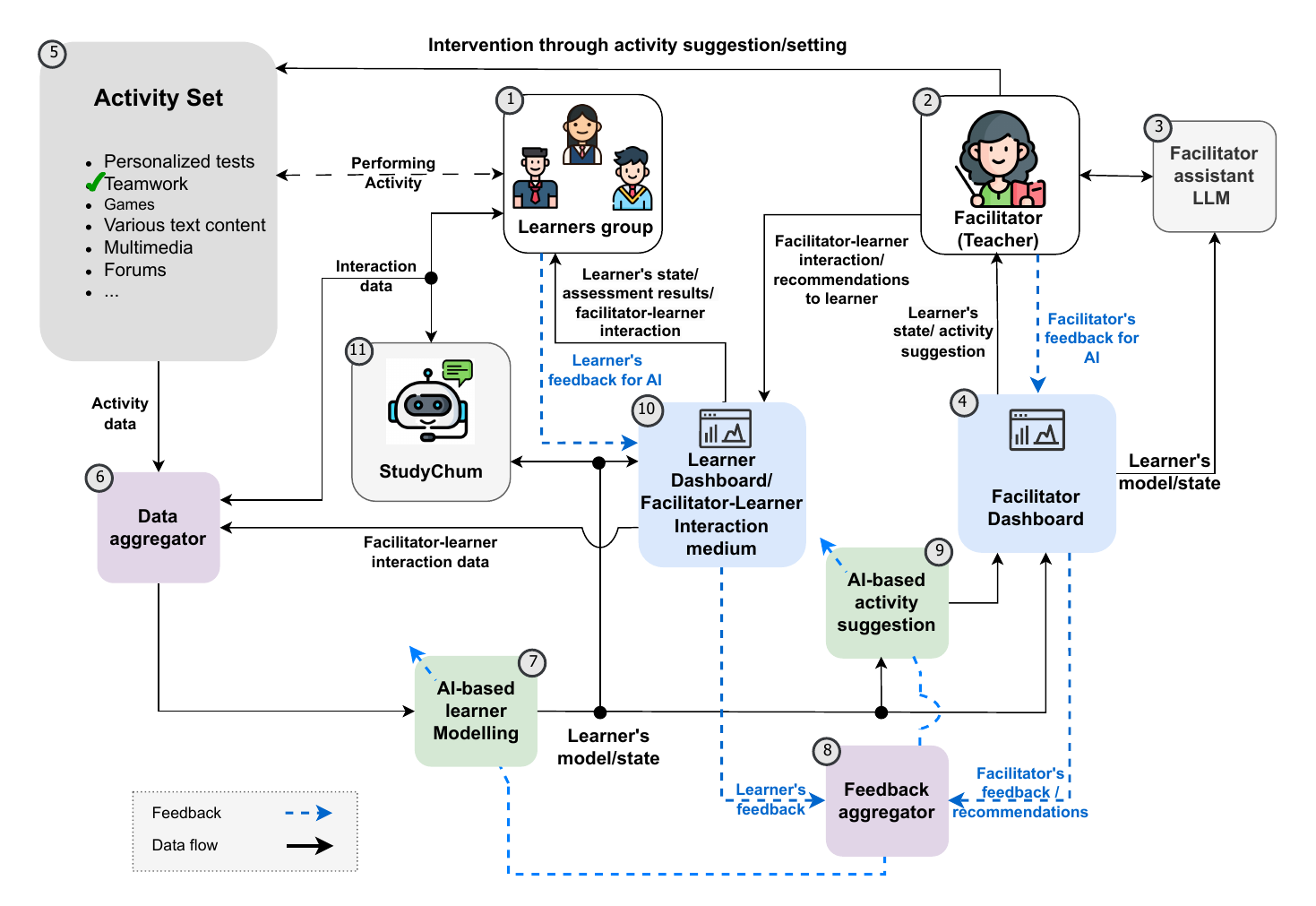}}
	\caption{
     The proposed framework. This figure illustrates a novel framework integrating our key principles with DL theory. Learners engage in group activities, each equipped with a personal dashboard for facilitator communication and self-monitoring. An AI agent, "StudyChum", proactively participates as a group member, dynamically adapting its role based on learner needs and models. Comprehensive data from learner interactions and actions feeds into AI-based learner modelling, refined through an Open Learner Model approach to minimize modelling errors. The system suggests personalized learning activities presented to learners after facilitator approval. Facilitators access a dashboard providing insights into learner progress and states, supported by a specialized LLM assistant for enhanced instructional decision-making.
 }
	\label{fig:conceptual}
\end{figure*}

To address these requirements, we suggest a digital framework that supports almost all digitally doable learning activities and provides high-resolution observability of learner activities required for multifaceted learner modelling. 
Our design aims to deliver scalable and low-cost quality education based on our proposed Digital Learning principles (see \Cref{table:Principles}). \Cref{fig:conceptual} illustrates our proposed framework, which comprises seven core components, namely learners group, AI-based learner modelling, AI-based activity suggestion, AI-based studyChum, AI-supported facilitator, and two dashboards for learners and facilitator.

\begin{table}[ht]
\centering
\caption{Our digital learning principles for designing the framework}
\vspace{-0.5em}
\renewcommand{\arraystretch}{1.5}
\resizebox{1\columnwidth}{!}{
\begin{tabular}[t]{m{1.8cm} m{5.7cm}}
\toprule
\textbf{Principle} & \textbf{Description} \\
\midrule
\textbf{Teacher in the loop} & Presence of teacher as facilitator in the closed-loop learning process with bidirectional feedback to learners \\
\textbf{AI-supported facilitator} & AI-enhanced facilitator support through learner modelling and learning activity suggestion system\\
\textbf{Learner-centred path} & AI-enabled learner-centred approach through giving choice and voice by offering personalized goals, activities, and learning paths \\
\textbf{Continuous learner modeling} & Activity-based continuous learner modeling using cognitive science-informed AI methods\\
\textbf{Collaboration} & Supporting collaborative learning for soft skill development and engagement improvement through teamwork and collaborative knowledge generation\\
\textbf{Personalized generative AI} & Enriching the learning environment using personalized LLM and generative AI integration in diverse roles as a personalized learning partner and teaching assistant\\
\textbf{Adaptive knowledge-based AI} & Learning science-driven and expert knowledge-based AI with continual learning based on human feedback and performance assessment\\
\textbf{Continuous assessment} & Integration of learning and continuous assessment\\
\bottomrule
\end{tabular}
}
\label{table:Principles}
\end{table}

Collaborative learning highly facilitates DL and is essential for soft skill development \cite{fullan2017deep, nadiyah2015development}. Therefore, we design a learners group as a component of our framework (component 1 in \Cref{fig:conceptual}) to facilitate collaboration among learners in which learners communicate in different modalities. For personalization purposes, collaborative activities require dynamically creating groups and monitoring learners' social behaviour, communications and contributions. As a result, collaborative learning activities impose extra complexity and highly increase the facilitator's load.  Our AI-based learner modelling and activity suggestion components take that load significantly. Therefore, the facilitators' load increment is not significant.  So, this system facilitates the "collaboration" and "continuous assessment" principles.

In this framework, grounded in DL, instead of being a mere instructor, the teacher becomes both a facilitator and a learners group member, working alongside learners (component 2 in \Cref{fig:conceptual}). This shift addresses "teacher in the loop" and "collaboration" principles. The facilitator has access to the learner's model and the suggested activities through her dashboard. Her duty here is to review and verify the suggested activities and to give feedback to the AI-based components accordingly. In addition, the facilitator has access to a learning science-informed LLM (component 3 in \Cref{fig:conceptual}) for consultation and assistance in performing her duties. This LLM is prompted by learner model data, suggested activities, and facilitator inquiries. This service partially compensates for limited access to professional facilitators, thereby supporting scalability.

Digital systems inherently offer high-resolution learner observability, which we leverage for comprehensive data collection. This data includes behavioural data, ethically collected physiological responses, reports, and learner interactions. This comprehensive data collection forms the foundation of most of our principles, making them achievable.

These comprehensive data allow learner modelling using cognitive science-informed machine learning (ML) and data science methods; refer to the learner modelling block as a component 7 in \Cref{fig:conceptual}. The output of modelling is the estimated learner's state/ model. To enhance transparency, trust, and learner engagement while mitigating potential AI-generated inaccuracies, we implement the Open Learner Model (OLM) approach \cite{winne2021open}. Learners review their predicted model via a personal dashboard, providing feedback that reduces errors in AI-driven modelling. We also incorporate facilitator feedback, enhancing the overall accuracy of the learner model. This combined approach leverages both AI predictions and human insights for more precise learner modelling.
This integrated system aligns with "AI-supported facilitator" and "continual learner modelling" principles, enabling scalable, cost-effective education. The key features, challenges, and potential solutions for AIEd are presented in \Cref{sec:AIEd_challenges}. Having the learner modelling component is crucial for quality education and supports economic and scalable systems by reducing the required teacher-to-learner ratio.

Personalization of activities is essential for DL. In our framework, the learner model is fed to an ML-based system (component 9 in \Cref{fig:conceptual}) that suggests some personalized learning activity from a given list; some learning activities are categorized in \Cref{table:Activities} \footnote{Details of each activity should be designed and customized for each topic. The details are culture-bounded and should match learners' demographical and cultural backgrounds.}. However, this component needs to improve itself over time and learn better suggestions for each learner or learners group. Therefore, this component uses the facilitator and learner feedback coming from their dashboards to improve itself. Then, suggestions of this component, along with the estimated learner state, are presented to the facilitator for review and approval.  Approved personalized activities and associated recommendations are then transmitted to the learner's dashboard. Also, the facilitator can intervene with these suggestions in the Activity Set.

Having learning science-informed AI-based learner modelling and activity suggestions significantly reduces the facilitator's decision-making load  and, therefore, is inline with having low-cost and scalable education. 
. In addition to supporting scalability, this load reduction provides the opportunity to increase the number and, consequently, the diversity of learning activities. Having diverse activities is crucial not only for personalization but also for learner modelling because each learner's characteristics, goals, etc, become visible in different learning activities. These diverse activities facilitate the realization of "AI-supported facilitator", "learner-centred path", and "continuous assessment" in educational practice. 

Integrating assessment into the learning process is crucial for closed-loop personalized learning \cite{delgashaly2018sengesh}. Traditional assessments, typically conducted infrequently, fail to capture the rich data available from learners' ongoing activities, which offer opportunities for high-frequency evaluation \cite{hussein2020pedagogical}. Moreover, traditional assessments are often unimodal, whereas learners demonstrate their proficiency levels, expertise, and potentials across various activities, enhancing their self-regulation. This multi-faceted approach is particularly significant for soft skills development.
Implementing a wide range of activities in conjunction with AI-based learner modelling provides the foundation for the "Continuous assessment" principle. This approach enables continuous evaluation across diverse learning activities, offering a more comprehensive and nuanced understanding of learner progress and needs.

Passive data collection may not fully capture all the characteristics and behavioural patterns of learners. According to Rauthmann et al. work \cite{rauthmann2014situational}, each learner demonstrates their unique and complex traits through interactions within a particular learning environment. This means that specific situations may not be reflected in the data, potentially leading to inaccurate insights into the learner's behaviour. We overcome this problem in two ways. The first one is a safe exploration of activity suggestions to the facilitator through component 9. 
The second way is to use a conversational personalized AI agent; we call it "StudyChum" in our framework (component 11 in \Cref{fig:conceptual}). StudyChum is a well-prompted \cite{bastani2024generative}, including learner state, LLM.

Actually, learners in different situations may require varying optimal interventions. At times, it may be beneficial for StudyChum to present itself at a level below the learners to enhance the learner's engagement by teaching StudyChum. Posing thought-provoking questions can stimulate the learner's curiosity and agency. Deliberately making mistakes for the learner to correct can strengthen their critical thinking skills. Alternatively, StudyChum may need to explain a topic based on the learner's interests and state as an assistant to enhance the learner's understanding \cite{kasneci2023chatgpt}. To achieve this adaptive role for StudyChum, adapting LLM's input prompt to each situation is crucial. Therefore, we suggest using automatic prompt engineering \cite{autoprompt} to learn the optimal prompts for each situation and based on the learner model as input. Also, these diverse interventions can encompass various engaging educational activities to foster the learner's growth and enhance their learning experience \cite{chen2020teaching}. 

The incorporation of StudyChum serves a dual purpose: it reduces the facilitator's workload \cite{kasneci2023chatgpt} while providing rich information for user modelling and assessment. Moreover, learner engagement with StudyChum constitutes a collaborative learning activity. Since we designed StudyChum as a group member and not as a source of knowledge, it facilitates the implementation of all of the principles except "Teacher in the loop." 

The facilitator dashboard as a component 4 in \Cref{fig:conceptual} aggregates and displays learner models, feedback, and AI model suggestions. This comprehensive view reduces the facilitator's workload by streamlining information access and saving time on administrative tasks. The dashboard enables educators to make informed decisions quickly and efficiently, enhancing their ability to provide personalized support.

Complementing the facilitator dashboard, the learner dashboard as a component 4 in \Cref{fig:conceptual} serves as a central hub for facilitator-learner collaboration. It captures interaction data and displays recommendations, feedback, and insights from AI and facilitators. This comprehensive approach facilitates ongoing communication and personalized support throughout the learning process.

\begin{table}[ht]
\centering
\caption{The list of proposed activities that facilitate DL}
\resizebox{1\columnwidth}{!}{
\begin{tabular}[t]{p{2.3cm} p{5.2cm}}
\toprule
\textbf{Category} & \textbf{Activities} \\
\midrule
\textbf{Active Learning and Exploration} & Studying/watching Content, Providing / Generating content, Problem-Solving, Project-based learning (PBL), Gamification, Doing real-world experiments, Search. \\[6pt]\\
\textbf{Assessment, Planning} & Self-report, Feedback, Exam, Planning, Asking Questions, Verification of LLMs.\\[6pt]\\
\textbf{Collaborative Interaction and Communication} & Group Work, Peer Assessment, Peer-teaching, Group discussion and debates, Brainstorming sessions. \\
\bottomrule
\end{tabular}
}
\label{table:Activities}
\end{table}


\section{AI Challenges in Framework Deployment and Practical Solutions} \label{sec:AIEd_challenges}

While AIEd offers promising potentials, its implementation in general faces numerous challenges \cite{chen2022two, zhai2021review, chiu2023systematic}. We believe that these difficulties mainly stem from the fact that most of the existing AI tools are not originally developed for human-in-the-loop applications in general and for educational applications in particular \cite{america2023doc}.
Our framework has been designed to address and reduce some of those challenges; however, there are still some difficulties in using AI in the framework.  Here, we discuss the challenges and suggest some solutions. We focus on using AI in components 3, 7, 9, and 11 (in \Cref{fig:conceptual}), where AI and LLMs have a central role in the scalability and affordability of quality education. 

StudyChum, a conversational assistant powered by LLMs, faces several challenges in delivering customized learning interactions. Safety is a primary concern, as StudyChum interacts with learners in educational contexts and provides proactive interventions, necessitating measures to address risks such as learner overreliance, potential bias, misinformation, and hallucinated suggestions \cite{kasneci2023chatgpt, kaddour2023challenges}. Personalization poses another hurdle, as LLMs lack inherent capabilities to generate tailored content or adapt to changing learner preferences and contexts. Scaling personalization for multiple learners simultaneously can be computationally intensive and costly \cite{kasneci2023chatgpt}. Providing personalized content may require detailed learner information, raising privacy concerns \cite{yan2024LLM, america2023doc}. Additionally, LLMs may struggle to fully grasp individual learner contexts and preferences across diverse user groups. StudyChum's potential use in various domains or lessons may also require specialized knowledge beyond the capabilities of a general-purpose LLM, further complicating its implementation and effectiveness.

To address StudyChum's challenges, we propose a multi-faceted approach. To mitigate overreliance, StudyChum is positioned as a peer learner, presents conflicting viewpoints, and encourages fact-checking \cite{tlili2023devil}. Safety concerns are addressed through a RAG system \cite{gao2023rag} incorporating established learning principles and ethical boundaries from cognitive science, learning science, and teaching expertise. Additionally, these concerns can be addressed with prompts designed to safeguard learning \cite{bastani2024generative}. Personalization is enhanced using Automatic Prompt Engineering with Reinforcement Learning (RL) \cite{autoprompt, li2024personalizedPrompt}, optimizing responses based on learner states and models. Also, StudyChum has the potential to adapt its level and role based on the learner through this approach. Data privacy is ensured through robust anonymization, encryption, transparent policies, and user control options. Bias is combated with continuous facilitator supervision and user feedback systems. To handle diverse learners and specialized content, we implement a comprehensive RAG system that incorporates learner-specific data, adaptive prompting techniques, dynamic learner profiles, and regular knowledge base updates. These solutions collectively aim to create a more effective, personalized, and ethically sound AI-powered learning assistant.

The facilitator-assisting LLM faces safety challenges common to other LLMs, with an additional critical requirement for explainability \cite{yan2024LLM}. The facilitator needs to understand AI's analysis of learner work and its recommendations, resources, and next steps.

To address this, the LLM should provide clear reasoning behind its decisions, enhancing transparency and building trust. This can be achieved by employing prompt structures informed by the science of learning. Additionally, implementing targeted training programs for educators is crucial to mitigate the risk of facilitator overreliance on the LLM \cite{kasneci2023chatgpt}. These strategies balance leveraging AI capabilities and maintaining human oversight in educational decision-making \cite{america2023doc}.

The Learner Modeling component faces critical challenges in explainability, processing heterogeneous data types, and large-scale inputs, which may be costly and noisy \cite{luan2020bigdata,wang2017heterogeneous}. Because its input data comes from diverse learning activities, facilitator-learner interaction, and StudyChum-learner interaction. Also, this component needs access to active data collection to adjust its modelling \cite{ouyang2021artificial}. Because AI models attempt to predict approximations of learner models, errors occur due to the dynamic, multi-faceted nature of human behaviour in educational contexts. Current AI models often focus narrowly on cognitive aspects, counting wrong answers and neglecting crucial elements such as self-regulation, engagement,  teamwork, and leadership skills. A more comprehensive approach should consider various learning aspects, including motivational and social elements. Additionally, AI models should adopt a strength-based approach, recognizing and building upon each learner's unique competencies, similar to how a one-to-one expert facilitator identifies and leverages learner assets. This asset-oriented approach would enhance AI models' equity and effectiveness in supporting diverse learners \cite{america2023doc}.
Additionally, adaptive models, like pure RL, face challenges with low convergence rates, limiting their practical application in dynamic educational environments.

We suggest using a neuro-fuzzy RL system \cite{shihabudheen2018fuzzy, derhami2008fuzzy} for learner modelling to ensure explainability. To address the challenges mentioned, we suggest incorporating effective features suggested by learning science research and local experts to create a more multifaceted and robust learner model. In addition, we propose implementing an expert-based dynamic attention mechanism that prioritizes relevant features to increase the accuracy of the modelling. This mechanism adapts continuously across diverse learners, improving data management and model accuracy. With all the measures mentioned, there is still no guarantee for sufficiently accurate learner modelling. Involving learners to check the critical aspects of their models at a low frequency, based on OLM \cite{winne2021open}, is our solution for improving the modelling accuracy and transparency. 
We propose initialising models with expert systems to mitigate the low convergence rates of RL models in dynamic educational environments. This reduces the need for fully random exploration and accelerates the learning process. This approach combines the adaptability of RL with the efficiency of established educational practices.

AI-based activity suggestion component faces two primary challenges: explainability and low convergence rates. Explainability is crucial for building trust, as educators and learners need to understand the rationale behind recommended activities \cite{america2023doc}. To address this, we can use a neuro-fuzzy system \cite{shihabudheen2018fuzzy, derhami2008fuzzy}, enhancing transparency and allowing facilitators to evaluate the AI's decisions effectively. Also, to prevent this component from irrelevant and inappropriate activity suggestions, we constrain it on the list of proposed activities (\Cref{table:Activities}). Low convergence rates hinder the model's ability to swiftly adapt to dynamic learner needs and educational contexts, potentially compromising the timeliness and relevance of activity suggestions. To mitigate this, we can start with the expert system as an activity suggestion component and then use hybrid approaches that combine pre-trained models with adaptive learning techniques, enabling faster adaptation while maintaining a foundation of established educational principles. These solutions aim to create a more transparent, trustworthy, and responsive AI-driven activity suggestion system.


\section{Summary and Discussion}

The global landscape of education is marred by stark inequalities, with access to quality learning opportunities varying widely across and within nations, irrespective of their economic development.
Socioeconomic barriers and rapid changes in the job market further exacerbate this issue, underscoring the urgent need for innovative solutions that can deliver quality education at scale and low cost. The DL theory emphasises the learner's central role in the learning path by targeting his/her agency and reduces the teachers' workload by redefining their role as facilitators, offering a promising solution for shaping a scalable learning environment. 

This paper outlined the various barriers to accessing quality education and proposed eight key principles, informed by learning science and AI, that are crucial for implementing an effective AIEd system for LLL. These principles serve as a guide for anyone looking to develop a learning science-informed AIEd system.

To realize these principles, we proposed a novel AI-powered digital framework. This framework, grounded in DL and AI principles within DLEs, leverages AI to enhance DLEs through AI-based learner modelling, personalized activity suggestions, and AI assistants for both facilitators and learners.

This approach aims to create personalized, collaborative, and engaging learning experiences. The framework is learner-centred while maintaining the teacher's presence, reducing their workload, and ensuring supervision. It enhances personalization in a collaborative manner, avoiding excessive individualism.

Additionally, the framework is designed to mitigate some existing challenges in AIEd. However, certain challenges persist. Processing heterogeneous input types remains a complex task. Accurately adjusting learner models based on feedback is challenging, as learners with low metacognition or self-confidence may provide inaccurate feedback. Therefore, a mechanism to validate and carefully apply this feedback to the model is necessary.

Furthermore, it's unclear whether LLMs can effectively understand and utilize high-level learner models and features to adjust their behaviour. If they cannot, Automatic Prompt Engineering becomes crucial for incorporating more detailed information in the prompt.

These remaining challenges highlight areas for future research and development in AIEd systems, particularly in refining learner models, improving feedback mechanisms, and enhancing LLM capabilities in educational contexts.

\bibliography{coling_latex}

\end{document}